\documentclass[11pt]{article}

\usepackage[margin=1in]{geometry}
\usepackage{graphicx}
\usepackage{booktabs}
\usepackage{amsmath}
\usepackage{amssymb}
\usepackage{caption}
\usepackage{float}
\usepackage[hidelinks]{hyperref}
\usepackage{xcolor}

\graphicspath{{figures/}}
\newcommand{\excess}{\texttt{excess\_acc}}
\newcommand{\alwaysup}{\texttt{always\_up}}

\title{\textbf{When Directional Accuracy Lies:\\ A Base-Rate-Honest Benchmark for LoRA-Adapted\\ TimesFM on Equity Forecasting}}
\author{Taizhen Cheung\\ \texttt{taizhen.cheung33@stu-mail.hunter.cuny.edu}}
\date{\today}

\begin{document}
\maketitle

\begin{abstract}
Large pretrained time-series models such as TimesFM are attractive starting points
for financial forecasting, but \emph{raw directional accuracy is a misleading
scoreboard in equity markets}. An early LoRA adapter in this project appeared to
reach roughly $80\%$ directional accuracy; we show this is not evidence of
forecasting skill. Over a long horizon in a rising market, a trivial ``always-up''
rule attains comparably high accuracy \emph{without using the input at all}. To
separate genuine skill from this base-rate artifact, we build a reproducible,
frozen-data benchmark with expanding \emph{walk-forward} train/validation/test
folds, a stratified \emph{held-out-ticker} split, honest baselines (zero-shot
TimesFM, always-up, random-walk, persistence, AR(1)), and paired significance tests
(McNemar, Diebold--Mariano) under Benjamini--Hochberg FDR control. We apply the
\emph{identical} method to two universes---a tech-heavy NASDAQ-100 and a broad
S\&P~500---and report the headline metric as \textbf{excess accuracy over the
always-up base rate}. Three findings emerge and replicate. First, when the
historical $\sim$80\% condition is recreated (a 2014+ bull window scored with raw
accuracy), the high number is a base rate of $\sim$0.70, and the fine-tuned model
scores \emph{below} it. Second, on the honest benchmark, pooled LoRA shows
\textbf{no directional skill over the base rate} at any horizon on either universe
(excess accuracy centered on zero, negative at the six-month horizon). Third, on the
pre-registered confirmatory test, \textbf{per-sector specialization is significantly
\emph{worse} than a single pooled adapter} (Diebold--Mariano $p<0.001$ on held-out
stocks at $h=128$). Fine-tuning's only measurable benefit is a statistically
significant reduction in point-forecast error relative to zero-shot TimesFM, which
nonetheless does not beat naive baselines and confers no tradeable directional edge.
The contribution is methodological: a defensible, fully seeded protocol that prevents
the base-rate trap, together with the replicated negative result it produces.
\end{abstract}

\section{Introduction}

Financial forecasting is an unusually treacherous setting for large time-series
models, because small apparent gains are easily confused with market drift, data
leakage, or evaluation artifacts. The problem is most acute for \emph{directional}
accuracy. A model that predicts ``up'' over a six-month equity horizon can look
highly accurate during a bull market even if it has learned nothing about the
relationship between past and future prices.

This project began with an apparent regression: a first TimesFM LoRA adapter
reported roughly $80\%$ directional accuracy, while later per-sector adapters dropped
toward $50\%$. Rather than treat the lower number as a failure, we treat the original
$80\%$ as the \emph{suspect} result. The flaw is that the early evaluation compared
raw accuracy against an implicit $50\%$ coin flip, even though long-horizon equity
direction is far from balanced: if most windows in the test period move up, an
always-up rule scores very well with zero skill.

The central question is therefore not ``can the model reach $80\%$?'' but
\emph{``does the model beat the right baseline on the same windows?''} We design and
implement a benchmark to answer this defensibly, and we run it on two universes with
the method held fixed, so that the conclusion does not hinge on a single dataset.
Our contributions are:
\begin{enumerate}
  \item A frozen, checksum-versioned dataset and a fully \textbf{seeded, reproducible}
  training$+$evaluation pipeline: every reported number traces to a committed
  \texttt{raw.json} and a \texttt{run\_meta.json} recording seed, code commit, clean-tree
  flag, and environment.
  \item A walk-forward train/validation/test protocol that never early-stops on the
  test window, plus a stratified held-out-ticker split for symbol generalization.
  \item The headline metric $\excess = \text{model accuracy} - \alwaysup\text{ accuracy}$,
  reported with block-bootstrap intervals and paired significance tests.
  \item A three-method, two-universe study (legacy pooled, walk-forward pooled, and
  per-sector adapters; NASDAQ-100 and S\&P~500) showing the negative result is robust
  and resolving the pre-registered per-sector-versus-pooled hypothesis.
\end{enumerate}

\section{Background and Related Work}

TimesFM~\cite{das2024timesfm} is a patched, decoder-only time-series foundation model
with internal reversible instance normalization (RevIN)~\cite{kim2021revin} and a
quantile output head. We adapt \texttt{google/timesfm-2.5-200m-pytorch} with
LoRA~\cite{hu2021lora}, which inserts a small number of trainable low-rank matrices
into selected linear layers while freezing the pretrained weights, making it cheap to
train specialized adapters.

The closest motivation is the financial fine-tuning study of Fu, Hirano, and
Imajo~\cite{fu2024financial}, which reports a \emph{modest} directional edge
($\approx 54\%$ at short horizons) rather than extreme accuracy. That reference is
exactly why an $80\%$ jump should not be accepted without checking base-rate
baselines, leakage, and out-of-sample generalization. We also draw on standard
forecast-evaluation practice: walk-forward validation, random-walk and persistence
baselines, MASE for point error~\cite{hyndman2006another}, quantile coverage and
pinball loss~\cite{gneiting2007strictly}, Diebold--Mariano tests~\cite{diebold1995comparing,harvey1997testing},
McNemar's paired test~\cite{mcnemar1947note}, the block bootstrap~\cite{kunsch1989jackknife},
and Benjamini--Hochberg FDR control~\cite{benjamini1995controlling}.

\section{Research Questions and Pre-registration}

\paragraph{RQ1 (primary, confirmatory).} Was the apparent directional accuracy a
base-rate artifact? Tested by $\excess = \text{acc} - \alwaysup\text{-acc}$ on
identical windows.

\paragraph{RQ2 (confirmatory).} Does per-sector specialization improve on a single
pooled adapter? Pre-registered as \emph{per-sector vs.\ pooled on held-out S\&P~500
stocks at $h=128$} (Diebold--Mariano).

\paragraph{RQ3.} Does LoRA fine-tuning improve on zero-shot TimesFM?

\paragraph{RQ4 (exploratory).} Does the conclusion depend on the universe? We
replicate the identical method on a tech-heavy NASDAQ-100 and a broad S\&P~500.

\noindent\textbf{Honesty note.} The per-sector test (RQ2) is reported on S\&P~500
only; NASDAQ-100 sector strata are too thin to train per-sector adapters (e.g.\ a
single financials name), so the universe comparison (RQ4) is restricted to the pooled
adapter and is treated as \emph{exploratory}, not as a registered hypothesis.

\section{Data}

Each universe is frozen once into a versioned, checksum-verified artifact and reused
by every run, so results never depend on a re-download (Table~\ref{tab:data}). Price
field is the split/dividend-adjusted close. The NASDAQ-100 membership and sector tags
are sourced from the advisor's \texttt{simofi} catalog and mapped onto a single sector
taxonomy shared with the S\&P~500.

\begin{table}[H]
\centering
\caption{Frozen datasets (2005-01-01 to 2026-01-01, daily).}
\label{tab:data}
\begin{tabular}{lcc}
\toprule
 & NASDAQ-100 & S\&P~500 \\
\midrule
Stocks ($+$ index/ETFs) & 100 ($+$QQQ) & 501 ($+$11 SPDR ETFs) \\
Sectors & 10 (tech-heavy: 41 tech) & 11 (balanced) \\
Seen / held-out stocks & 76 / 24 & 394 / 107 \\
Sectors with a held-out set & 7 / 10 & 11 / 11 \\
\bottomrule
\end{tabular}
\end{table}

\paragraph{Limitation (survivorship).} Both universes are \emph{current-membership
snapshots}, not point-in-time constituents. This inflates the up base rate and the
long-horizon upward drift, so results are a benchmark over a frozen current-constituent
universe, not a live investable simulation (Section~\ref{sec:limits}).

\section{Methodology}

\subsection{Walk-forward folds}
Evaluation uses three expanding folds; validation is the \emph{only} signal for early
stopping and model selection, and the test window is never used for selection
(Table~\ref{tab:folds}).

\begin{table}[H]
\centering
\caption{Expanding walk-forward folds (train $\rightarrow$ validation $\rightarrow$ test).}
\label{tab:folds}
\begin{tabular}{cccc}
\toprule
Fold & Train & Validation & Test \\
\midrule
0 & 2005--2019 & 2019--2020 & 2020--2022 \\
1 & 2005--2021 & 2021--2022 & 2022--2024 \\
2 & 2005--2023 & 2023--2024 & 2024--2026 \\
\bottomrule
\end{tabular}
\end{table}

\subsection{Held-out tickers}
Within each sector, stocks are split $\sim$80/20 into \emph{seen} and \emph{held-out}
with a fixed seed (42), \emph{fixed across folds}, so held-out names are never trained
in any fold. ETFs are excluded from the holdout pool and reported separately. Every
method is scored on both splits.

\subsection{Normalization (leakage control)}
A per-series log transform followed by $z$-scoring is fit only on \emph{pre-target
history}: the train period for validation windows, train$+$validation for test
windows. No target-window observation is used. Direction is invariant to this monotone
transform; returns and point error use inverse-transformed prices.

\subsection{Reproducibility}
Training is fully seeded (seed $=42$ across Python/NumPy/torch/CUDA, with a seeded
DataLoader generator; the seed is set \emph{before} LoRA initialization). The adapter
scored is the \emph{best-validation} checkpoint, not the last epoch. Each run writes a
\texttt{run\_meta.json} manifest recording seed, git commit, a tracked-files-only
clean-tree flag, GPU, library versions, the full dependency freeze, hyperparameters,
and the per-adapter best epoch. All runs in this paper were produced at one code commit
on a clean tree, seed 42, on an NVIDIA A100 (torch 2.11.0$+$cu128, Python 3.12); a
two-run reproducibility gate confirmed bit-identical Table-A accuracies. We do not
enable fully deterministic CUDA kernels (they raise on TimesFM operators), so results
are single-seed deterministic up to negligible GPU-atomic jitter; multi-seed robustness
is left to future work.

\subsection{Models and baselines}
\textbf{Zero-shot TimesFM} is the pretrained base with no adapter, served from a
dedicated instance with an assertion that adapter loading cannot mutate it.
\textbf{Pooled LoRA} is one adapter trained on all seen stocks. \textbf{Per-sector
LoRA} trains one adapter per GICS sector (seen stocks of that sector plus its SPDR ETF
anchor) and routes each ticker to its sector's adapter at evaluation. Naive baselines
are \alwaysup{} (predict up every window---the base rate), random-walk (last value; no
directional call), persistence (repeat last return sign), and AR(1) (direct $h$-period
return). All adapters share the configuration in Table~\ref{tab:hp}; only the data
differ across universes and methods.

\begin{table}[H]
\centering
\caption{Training configuration (identical across universes and methods).}
\label{tab:hp}
\begin{tabular}{llll}
\toprule
Base & timesfm-2.5-200m & LoRA rank / $\alpha$ / dropout & 32 / 64 / 0.05 \\
Context / horizon & 512 / 128 & Batch & 512 \\
Optimizer & AdamW (lr $10^{-4}$, wd 0.01) & Schedule & cosine warm restarts \\
Grad clip & 1.0 & Loss & MSE $+\,0.3\cdot$directional \\
Max epochs / patience / min & 80 / 15 / 30 & Seed & 42 \\
\bottomrule
\end{tabular}
\end{table}

\subsection{Metrics}
The headline directional metric is $\excess = \text{acc} - \alwaysup\text{-acc}$ on
identical windows; we also report balanced accuracy and Matthews correlation. Point
error (price space, per-ticker then macro-averaged) uses MAE, RMSE, sMAPE, and MASE.
Calibration uses empirical quantile coverage versus nominal and pinball loss. Economic
metrics use real per-trade returns on non-overlapping $h$-day windows; maximum drawdown
is computed on a unit-stake equity curve with per-trade returns floored at $-100\%$ (a
margin stop), so a short squeeze cannot drive it below $-1$.

\subsection{Significance}
Excess accuracy versus always-up uses the paired McNemar test plus a block-bootstrap
interval; forecast-error comparisons use Diebold--Mariano on normalized-space errors
with a sample-size-capped Newey--West variance (negative $\Rightarrow$ first method has
lower loss). The exploratory family is FDR-controlled by Benjamini--Hochberg.

\section{Results}

All headline numbers are on \emph{held-out stocks}, averaged over the three
walk-forward folds. Figures are produced directly from the committed result files.

\subsection{Recreating the legacy $\sim$80\% as a base-rate artifact (RQ1)}
We first recreate the historical condition that produced the $\sim$80\% claim: a pooled
adapter trained on a 2014+ bull window (train 2014--2023, early-stop 2023--2024) and
scored with raw accuracy on the 2024--2026 test window. The recreated base rate is
high but the model lands \emph{below} it (Table~\ref{tab:legacy}): at $h=128$ the
always-up rule alone scores $0.704$ on held-out stocks, while the fine-tuned model
scores $0.626$ ($\excess = -0.08$). The original $0.778$ figure was computed on only
$\sim$45 windows and does not reproduce on the full universe; honestly measured, the
``high accuracy'' is a base rate of $\sim$0.70 that the model does not beat. The
mechanism behind the $80\%$ claim---high raw accuracy that looks impressive against a
$50\%$ reference---is therefore confirmed, and shown to be the trend rather than skill.

\begin{table}[H]
\centering
\caption{Legacy 2014+ condition, held-out stocks: the recreated base rate is high, but
the model scores below it (no excess skill).}
\label{tab:legacy}
\begin{tabular}{lcccc}
\toprule
Horizon $h$ & 8 & 32 & 64 & 128 \\
\midrule
\alwaysup{} base rate & 0.684 & 0.629 & 0.636 & 0.704 \\
Model raw accuracy    & 0.559 & 0.526 & 0.530 & 0.626 \\
Excess (model $-$ base rate) & $-0.125$ & $-0.103$ & $-0.106$ & $-0.078$ \\
\bottomrule
\end{tabular}
\end{table}

\subsection{No directional skill over the base rate (RQ1, both universes)}
\begin{figure}[H]
\centering
\includegraphics[width=0.86\textwidth]{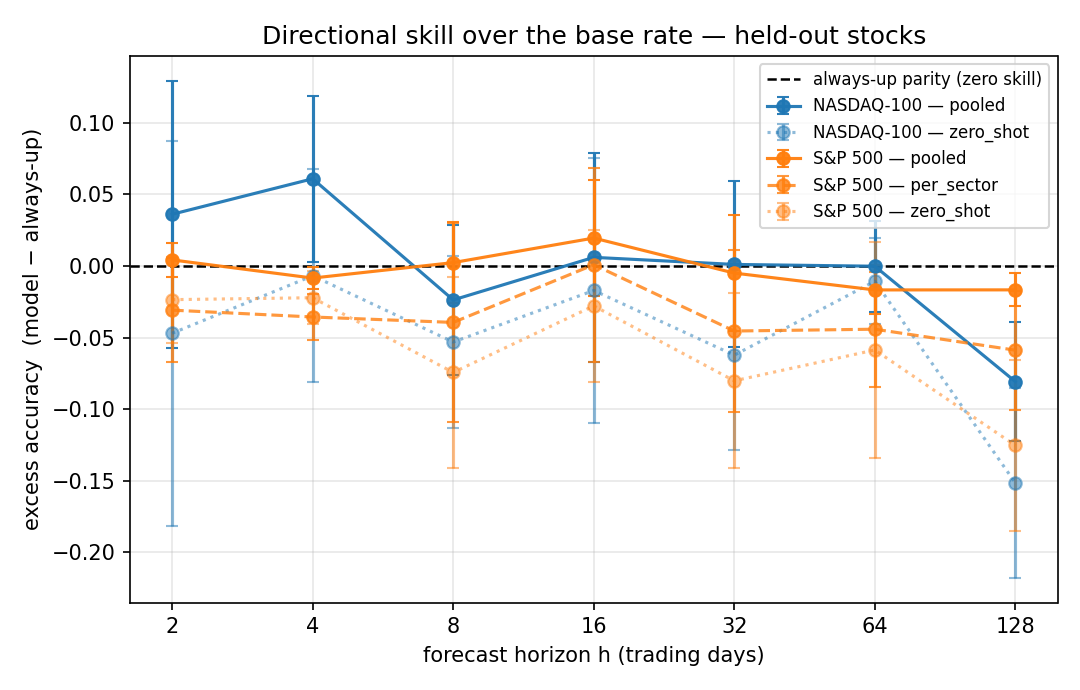}
\caption{Excess directional accuracy versus horizon, held-out stocks, both universes.
The dashed line is always-up parity (zero skill). Pooled curves hug the line with
intervals spanning zero; zero-shot sits consistently below it.}
\label{fig:excess}
\end{figure}

On the honest walk-forward benchmark, raw accuracy is high at the long horizon
($0.63$--$0.71$) but tracks the always-up base rate; excess accuracy is centered on
zero with bootstrap intervals spanning zero at every horizon, and is negative at
$h=128$ (Table~\ref{tab:excess}, Figure~\ref{fig:excess}). Zero-shot TimesFM is
\emph{below} the base rate everywhere. At $h=128$ the base rate alone is $0.710$
(NASDAQ) and $0.658$ (S\&P), above pooled accuracy of $0.630$ and $0.641$: no
configuration beats the base rate at the long horizon on either universe.

\begin{table}[H]
\centering
\caption{Excess accuracy (model $-$ \alwaysup{}), held-out stocks, by horizon.
Positive means skill over the base rate; intervals span zero throughout.}
\label{tab:excess}
\begin{tabular}{lccccccc}
\toprule
$h$ & 2 & 4 & 8 & 16 & 32 & 64 & 128 \\
\midrule
\multicolumn{8}{l}{\textit{NASDAQ-100}}\\
\quad pooled    & $+0.036$ & $+0.061$ & $-0.024$ & $+0.006$ & $+0.001$ & $-0.000$ & $-0.081$ \\
\quad zero-shot & $-0.047$ & $-0.007$ & $-0.053$ & $-0.017$ & $-0.062$ & $-0.011$ & $-0.151$ \\
\multicolumn{8}{l}{\textit{S\&P~500}}\\
\quad pooled     & $+0.004$ & $-0.009$ & $+0.002$ & $+0.019$ & $-0.005$ & $-0.017$ & $-0.017$ \\
\quad per-sector & $-0.031$ & $-0.036$ & $-0.039$ & $+0.001$ & $-0.045$ & $-0.044$ & $-0.059$ \\
\quad zero-shot  & $-0.024$ & $-0.022$ & $-0.074$ & $-0.028$ & $-0.080$ & $-0.059$ & $-0.125$ \\
\bottomrule
\end{tabular}
\end{table}

\begin{figure}[H]
\centering
\includegraphics[width=0.92\textwidth]{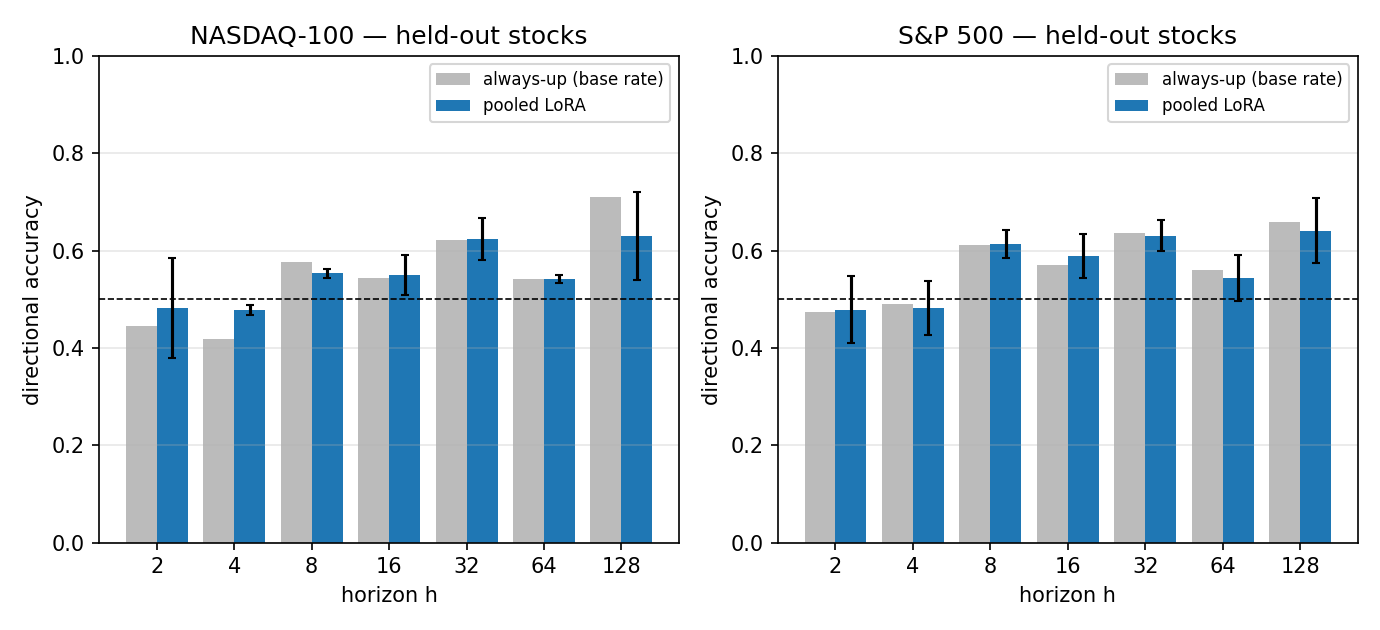}
\caption{Raw directional accuracy of pooled LoRA tracks the always-up base rate at
every horizon; at $h=128$ the base rate is higher. High raw accuracy is not skill.}
\label{fig:baserate}
\end{figure}

\subsection{Per-sector versus pooled: the pre-registered test (RQ2)}
The registered confirmatory comparison resolves cleanly against specialization. On
held-out S\&P~500 stocks, per-sector accuracy ($0.599$ at $h=128$) is below pooled
($0.641$) and below the base rate ($0.658$), and the Diebold--Mariano test rejects
equality \emph{in pooling's favor}: per-sector has significantly higher forecast loss
at the long horizon (Table~\ref{tab:persector}). Sector specialization therefore
\emph{hurts} relative to a single pooled adapter---most plausibly because each sector
adapter trains on far less data---and neither beats the base rate.

\begin{table}[H]
\centering
\caption{Pre-registered test: per-sector vs.\ pooled, held-out S\&P~500 stocks,
Diebold--Mariano statistic per fold (positive $\Rightarrow$ per-sector has higher
loss, i.e.\ \emph{worse}). Significant in pooling's favor at the long horizon.}
\label{tab:persector}
\begin{tabular}{lccc}
\toprule
Horizon & Fold 0 & Fold 1 & Fold 2 \\
\midrule
$h=8$   & $+3.26$ ($p{=}.001$) & $+2.15$ ($p{=}.031$) & $+3.57$ ($p{<}.001$) \\
$h=32$  & $+3.43$ ($p{=}.001$) & $+1.28$ ($p{=}.20$)  & $+2.20$ ($p{=}.028$) \\
$h=128$ & $+5.12$ ($p{<}.001$) & $+1.03$ ($p{=}.30$)  & $+5.50$ ($p{<}.001$) \\
\bottomrule
\end{tabular}
\end{table}

\subsection{Fine-tuning versus zero-shot: point error only (RQ3)}
\begin{figure}[H]
\centering
\includegraphics[width=0.62\textwidth]{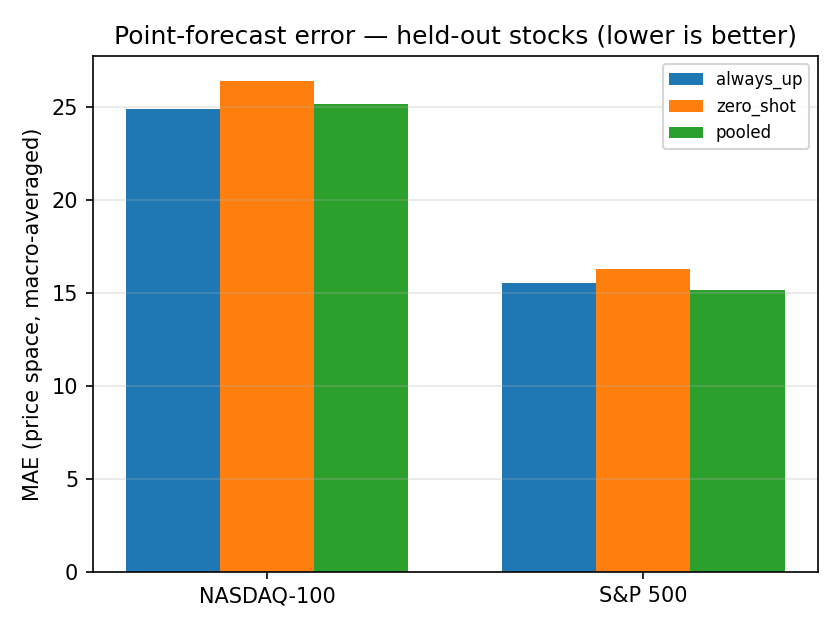}
\caption{Point-forecast MAE on held-out stocks: pooled LoRA is below zero-shot but ties
the always-up baseline.}
\label{fig:mae}
\end{figure}

Pooled LoRA has \emph{lower} point error than zero-shot TimesFM on both universes
(held-out MAE $15.15$ vs.\ $16.31$ on S\&P; $25.15$ vs.\ $26.44$ on NASDAQ;
Table~\ref{tab:mae}, Figure~\ref{fig:mae}), and Diebold--Mariano rejects equality in
pooled's favor in a majority of cells after FDR (15/21 S\&P-seen, 12/21 S\&P-held-out;
7/21 and 4/21 on NASDAQ). However, pooled does \emph{not} beat the naive always-up
point forecast ($15.56$ on S\&P, $24.91$ on NASDAQ), and the directional and economic
edge is absent. Fine-tuning sharpens the point forecast without conferring tradeable
skill---consistent with the observation that the best-validation epoch is reached
within a few epochs (S\&P: $\{1,2,1\}$; NASDAQ: $\{5,16,3\}$ of 44 run), after which
validation loss rises.

\begin{table}[H]
\centering
\caption{Point-forecast error (MAE, held-out stocks). LoRA $<$ zero-shot, but ties the
naive baseline. MAE is not comparable across universes (different price levels).}
\label{tab:mae}
\begin{tabular}{lccc}
\toprule
 & pooled & zero-shot & \alwaysup{} \\
\midrule
NASDAQ-100 & 25.15 & 26.44 & 24.91 \\
S\&P~500   & 15.15 & 16.31 & 15.56 \\
\bottomrule
\end{tabular}
\end{table}

\subsection{Calibration and statistical power}
\begin{figure}[H]
\centering
\includegraphics[width=0.92\textwidth]{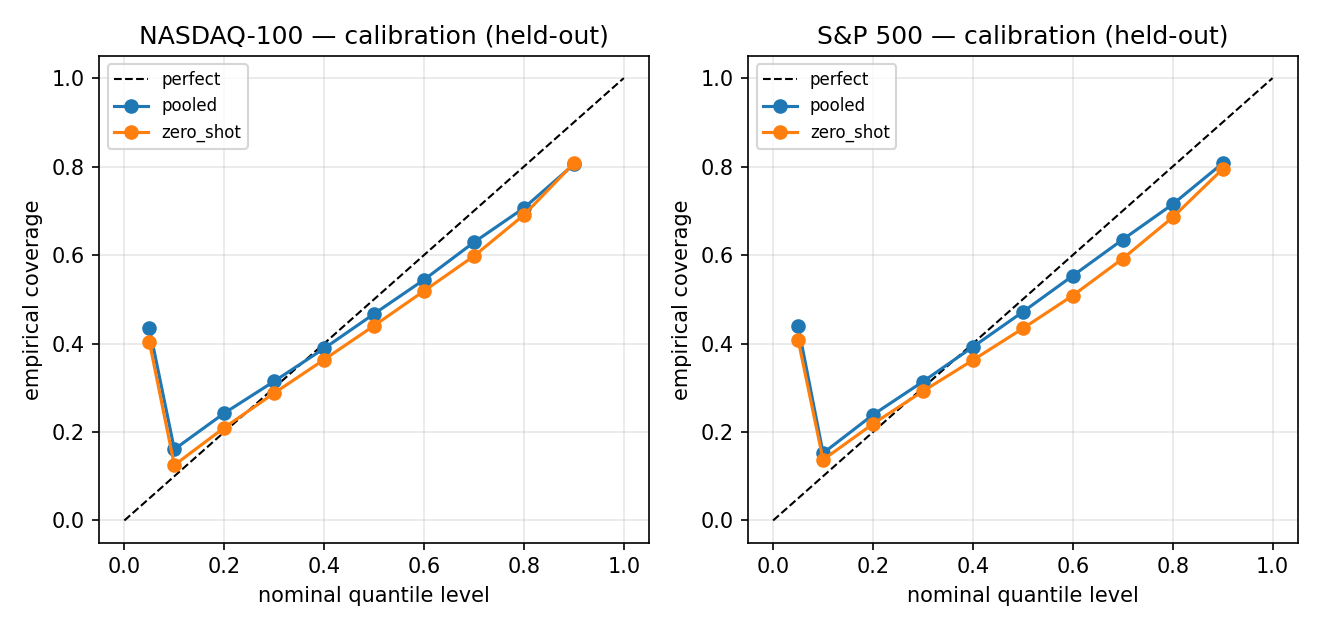}
\caption{Quantile reliability (held-out stocks): empirical coverage versus nominal for
pooled and zero-shot. Both are poorly calibrated in the tails on both universes.}
\label{fig:calib}
\end{figure}

The quantile tails are poorly calibrated for both models on both universes: the
nominal-$0.05$ quantile has $\sim$0.44 empirical coverage (pooled $0.440$ / zero-shot
$0.408$ on S\&P; $0.436$ / $0.404$ on NASDAQ), a base-model trait LoRA does not fix
(Figure~\ref{fig:calib}). The long horizon is severely underpowered: at $h=128$ the
held-out McNemar discordant-pair count per fold is only $\{5,5,20\}$ (S\&P) and
$\{3,10,10\}$ (NASDAQ), so we make \emph{no} significance claim at $h=128$ for the
directional tests; it is reported for completeness.

\subsection{Universe replication (RQ4)}
The two universes differ in size (100 vs.\ $\sim$500 names), sector mix (41\% vs.\
$\sim$15\% technology), and price level, yet the qualitative result is identical:
excess accuracy $\approx 0$, negative at $h=128$, zero-shot below the base rate, and
LoRA better than zero-shot on point error only. We therefore do \emph{not} claim one
universe is ``better'' (the comparison is confounded); the value of the replication is
that the negative finding is robust to the universe rather than an artifact of one
dataset.

\section{Discussion}

The interpretive risk this project set out to avoid---mistaking raw directional
accuracy for skill---is exactly the trap the original $80\%$ fell into. With an honest
base-rate baseline on identical windows, excess accuracy collapses to zero and is
negative at the six-month horizon: the equity uptrend, not the model, produced the high
accuracy, and this holds on a tech-heavy and a broad universe alike. The pre-registered
comparison adds a second, sharper negative: sector specialization is significantly worse
than pooling, so the intuitive ``train a specialist per sector'' recipe is
counterproductive here, most plausibly because each adapter sees far less data.

The one genuine, statistically supported effect is that LoRA reduces point-forecast
error relative to zero-shot TimesFM. This is a real but narrow benefit: it does not
beat naive baselines, does not produce directional accuracy over the base rate, and
does not yield a positive risk-adjusted return. Combined with the early best-validation
epoch, the picture is of a base model already near its useful operating point on these
series, with LoRA offering a small refinement of magnitude but not of direction. For
practice the lesson is methodological: report excess-over-base-rate, hold out tickers,
walk forward, and test paired significance before claiming a foundation model ``works''
on equities.

\section{Limitations}
\label{sec:limits}
\begin{enumerate}
  \item \textbf{Long-horizon power.} Non-overlapping $h=128$ windows yield only a few
  hundred (S\&P) or few dozen (NASDAQ) windows per fold and $\{5,5,20\}$/$\{3,10,10\}$
  McNemar discordant pairs; $h=128$ directional significance is not estimable and is not
  claimed.
  \item \textbf{Survivorship bias.} Current-membership snapshots inflate the up base
  rate and long-horizon drift; results do not transfer unchanged to a point-in-time
  universe.
  \item \textbf{Universe comparison is confounded.} NASDAQ vs.\ S\&P differ in size,
  sector mix, and price level, so no ``which universe is better'' claim is made.
  \item \textbf{Single seed.} Results are single-seed deterministic, not multi-seed
  robust.
  \item \textbf{Windowing.} Non-overlapping windows (stride $=$ horizon) under-sample
  short horizons.
  \item \textbf{Scope.} Equities and sector ETFs only; no forex or macro; CRPS is an
  even-grid quantile approximation.
\end{enumerate}
Future work: multi-seed mean$\pm$SD; a matched 100-stock S\&P subsample to deconfound
universe size from sector mix; point-in-time membership to remove survivorship; and
overlapping windows or longer test spans to recover long-horizon power.

\section{Conclusion}

Across two equity universes, with an identical seeded, reproducible method,
LoRA-adapted TimesFM shows no directional skill over the always-up base rate at any
horizon, zero-shot TimesFM is worse than the base rate, and per-sector specialization
is significantly worse than a single pooled adapter. Fine-tuning's only measurable
benefit is lower point-forecast error than zero-shot, which beats neither the naive
baselines nor the base rate. The apparent $\sim$80\% directional accuracy that
motivated the project was a base-rate/trend artifact---recreated here as a $\sim$0.70
base rate the model fails to beat. The deliverable is a benchmark protocol that makes
such artifacts visible, and the replicated, honest negative result it produces.

\section*{Acknowledgements}

The author thanks Prof.~S.~A.~Kwon for advising this project, for early
exploratory work on sector classifications, for providing the \texttt{simofi}
membership and sector catalog used to construct the NASDAQ-100 universe, and
for feedback on the manuscript. The author also thanks Prof.~Indranil
SenGupta for reviewing the manuscript.

\end{document}